# A DECISION SUPPORT SYSTEM FOR COMMUNICATION NETWORK ANALYSIS BASED ON HYBRID PETRI NET MODELS

Simona Iuliana Caramihai, Călin Munteanu, Janetta Culiță

„Politehnica" University of Bucharest, Romania
Dept. of Automatic Control and Computer Science
E-mail: sic@ics.pub.ro, mc_aurel@yahoo.com, jculita@yahoo.com,

**Abstract:** It is well known that the complex system operation requires the use of new scientific tools and computer simulation. This paper presents a modular approach for modeling and analysis of the complex systems (in communication or transport systems area) using Hybrid Petri nets. The performance evaluation of the hybrid model is made by a simulation methodology that allows building up various functioning scenarios. A Decision Support System based on the above mentioned methodologies of modeling and analysis will be designed for performance evaluation and time optimization of large scale communication and/or transport systems.

**Keywords:** hybrid Petri nets, Decision Support Systems, performance evaluation, simulation.

## 1. INTRODUCTION

Complex systems represent a fast growing area of interest for control and optimization. Domains as communication and transport are important fields of interest from this point of view, since the present trend consists in building aggregated networks either for informatics or for transport (such as the European unified railway system project).

Performance evaluation, analysis and optimization of these networks necessitate particular methodologies and formal instruments. Since pure mathematical models are not easy to handle for these categories of systems, decision is often taken on the basis of the simulation of various functioning scenarios.

Decision Support Systems (DSS) ensuring the appropriate framework for modeling, analysis and comparison of those scenarios are valuable tools for network managers.

The paper proposes a DSS architecture for complex network analysis, based on hybrid Petri Nets modeling and analysis. Petri Nets (PN) were defined in 1962 by Karl Adam Petri, for modeling and analysis of information systems. Their modeling power, capacity of representing concepts as parallelism, synchronization, resource sharing a.s.o. in a clear, intuitive graphical format, have contributed to their wide development and utilization, for very different domains.

More recently, *continuous* PNs (CPN) were defined in [3]. Autonomous continuous PNs have been shown to be a limit case of discrete PNs. In a continuous PN, the markings, arc weights and firing quantities are non-negative, as in a discrete PN, but are *not necessarily integers*. In a *timed continuous* PN (TCPN), maximal speeds are associated with transitions. Other authors have developed interesting results concerning these types of PNs ([5], [8], [9]).

The initial motivation leading to the concept of continuous PN was an approximate modeling of discrete systems with a large number of states, as a consequence of the management of a large number of entities by the modeled process. By this approach it was possible to analyze the flow of entities instead of following the evolution of every one. Another domain of application of this type of nets is for continuous system modeling.

*Hybrid* PNs (HPN) contain both a discrete and a continuous part, usually interacting ([6], [4]). In a timed hybrid PN (THPN) timings are associated with discrete transitions and maximal speeds are associated with continuous transitions. The instantaneous behavior is analyzed by the following way: a stable marking of the discrete part is sought, then the instantaneous firing speeds of the continuous transitions are calculated.

The semantics related to instantaneous firing speeds (i.e. local calculation when given the marking and feeding flows of the input continuous places of a continuous transition) is relatively easy to define. However, automatic calculation for the global CPN is difficult. Iterative algorithms presented in [3] and [1] do not work in all cases. Calculation by resolution of a linear programming problem (LPP) was used for some specific cases in [2].

The bases of a speed calculation method for CPN are presented in [7]. Many more details are given in [4] and software for analysis of Hybrid and Continuous Petri nets is available at http://sirphyco.lag.ensieg.inpg.fr/.

The paper will use the Hybrid Petri Nets formalism as the basic approach for designing a Decision Support System for communication and/or transport networks. The next section will present main issues on HPN formalism. Section 3 describes the specifications of the network analysis problem and the consequent structure and operation of the DSS. Section 4 will present a short case study, illustrating the methodology of analysis implemented by the DSS. Finally, the conclusion section will present some future research directions

## 2. HYBRID PETRI NETS FORMALISM

Before the HPN formalism is introduced, some definitions relative to the Continuous Petri Net (CPN) features are presented in the sequel. It is assumed that the basic formalism of discrete Petri Nets and timed discrete Petri Nets (TPN) is well known.

As in [4] a *marked autonomous* CPN is a 5-tuple $R = \langle P, T, \text{Pre}, \text{Post}, \mathbf{m}_0 \rangle$ such that

$P = \{P_1, P_2, \ldots, P_n\}$ is a finite, not empty, set of *continuous* places;

$T = \{T_1, T_2, \ldots, T_m\}$ is a finite, not empty, set of transitions;

$P \cap T = \varnothing$, the sets $P$ and $T$ are disjointed;

$\text{Pre} : P \times T \to Q_+$ is the input incidence application; $\text{Pre}(P_i, T_j)$ is the weight of the arc $P_i \to T_j$

$\text{Post} : P \times T \to Q_+$ is the output incidence application; $\text{Post}(P_i, T_j)$ is the weight of the arc $T_j \to P_i$

$\mathbf{m}_0 : P \to R_+$ is the vector associated with the initial marking;

The main features specific to autonomous CPNs are: the existence of continuous places and transitions, the real positive value of a marking that could change continuously in a place; the possibility of multiple firing of an enabled transition (*at one go*- meaning more times simultaneously firing of a transition). For these reasons CPNs are particularly suitable for flow modeling.

The *enabling degree* of transition $T_j$ for marking $\mathbf{m}$, denoted by $q$ or $q(T_j, \mathbf{m})$ is the real number $q$ such that

$$q = \min_{i:P_i \in {}^\circ T_j} \left( \frac{m(P_i)}{\text{Pre}(P_i, T_j)} \right) \quad (1)$$

If $q > 0$, transition $T_j$ is $q$-*enabled*. However, a transition can fire with a firing quantity less than its enabling degree.

The reachability property is defined in terms of macro-markings. A *macro-marking* $\mathbf{m}_j^*$ is the union of all markings $\mathbf{m}_k$ with the same set $P^+(\mathbf{m}_k)$ of marked places, where $P^+(\mathbf{m}_k) = \{P_i | \mathbf{m}_k(P_i) > 0\}$. In order to obtain the reachability graph, it is sufficient to know the macro-markings. Given a CPN, two types of events can cause a change of macro-marking in a reachability graph
- *C1-event*: the *marking* of a marked place *becomes zero*.
- *C2-event:* an unmarked place *becomes marked*.

By associating a maximal speed to each transition a timed continuous PN (TCPN) is obtained. The maximal speed could be constant or time dependent. Formally, a TCPN is defined by a pair
$$TCPN = \langle R, V \rangle, \text{ where}$$
$R$ is a a marked autonomous CPN;
$V$ is a function from the set $T$ of transitions to $Q_+ \cup \{\infty\}$ that associates to every transition $T_j$ a *maximal speed* $V_j$, $V: T \to Q_+ \cup \{\infty\}$, $V(T_j) = V_j$.

Although a transition is initially associated with a maximal speed $V_j$, during the evolution of a CPN it can be characterized by an *instantaneous speed* $v_j$,

$$v_j = \begin{cases} V_j, & m(°T_j) > 0 \\ 0, & m(°T_j) = 0 \end{cases} \quad (2)$$

A *transition* $T_j$ in a TCPN is *non-immediate* if it is characterized by a finite maximal firing speed. A non-immediate transition is *enabled* at $t$ if $\tilde{m}_i(t) > 0$ for every $P_i \in °T_j$, where $\tilde{m}_i$ represents the marking used for the calculation of the speed vector. It is *strongly enabled* if $m_i(t) > 0$ for every $P_i \in °T_j$, and *weakly enabled* otherwise.

The behaviour of a TCPN is analysed by means of the evolution graph. This is algorithmically computed and it is composed of phases. A phase represents a time interval of linear evolution of TCPN during which the instantaneous speed vector remains constant. The computation of the instantaneous speed vector is based on the following concepts
- the *feeding speed* of a place $P_i$ in a TCPN is

$$I_i = \sum_{T_j \in °P_i} \text{Post}(P_i, T_j) \cdot v_j \quad (3)$$

If $I_i > 0$, place $P_i$ is said to be fed.
- the *draining speed* of a place $P_i$ in a TCPN is

$$O_i = \sum_{T_j \in P_i°} \text{Pre}(P_i, T_j) \cdot v_j \quad (4)$$

- the *balance* of the marking of a place $P_i$ in a TCPN is

$$B_i = I_i - O_i \quad (5)$$

The balance corresponds to the time derivative of its marking, $\dfrac{dm_i(t)}{dt} = B_i(t)$, and $m_i(t+dt) = m_i(t) + B_i(t) \cdot dt$

If a transition $T_j$ is *strongly enabled* it is fired at its maximal speed $v_j(t) = V_j$. Given the set of places $Q_i(t) = \{P_i | P_i \in °T_j, m_j(t) = 0\}$, for a *weakly enabled* transition $T_j$ not involved in an actual conflict related to a place in $Q_i(t)$, the instantaneous firing speed of $T_j$ is

$$v_j(t) = \min_{P_i \in Q_j(t)} \left( \frac{1}{\text{Pre}(P_i, T_j)} \sum_{T_j \in °P_i} \text{Post}(P_i, T_k) \cdot v_k(t), V_j \right)$$
(6)

For a marking vector $\mathbf{m}(t)$ at the moment $t$, and having specified the feeding speeds $I_i(t)$ of each place $P_i$ involved in a structural conflict, an *effective conflict* exists if
1) $m_i(t) = 0$,

2) $I_i(t) < \sum_{T_j \in P_i°} \min_{\{P_h / P_h \in °T_j \& m_h(t)=0\}} \left( \frac{1}{\text{Pre}(P_h, T_j)} \sum_{T_k \in °P_h} \text{Post}(P_h, T_k) \cdot v_k, V_j \right)$

The value of the feeding speed of a null marking place $P_i$ (involved in a structural conflict) is not sufficient for assuring the firing of every output transition $T_j \in P_i°$ at their firing speed computed in the lack of conflict.

If there are effective conflicts in the TCPN, a deterministic evolution of TCPN is obtained when certain rules for conflict solving are applied: 1. priority rules between the conflict transitions 2. sharing.
A marked *autonomous hybrid Petri Net* (HPN) is formally described by a sextuple

$R = \langle P, T, \text{Pre}, \text{Post}, \mathbf{m}_0, h \rangle$ where:

$P = \{P_1, P_2, \ldots, P_n\}$ is a finite, not empty, set of places;

$T = \{T_1, T_2, \ldots, T_m\}$ is a finite, not empty, set of transitions;

$P \cap T = \varnothing$, the sets $P$ and $T$ are disjointed;

$h: P \cup T \rightarrow \{D, C\}$, called "hybrid function", indicates for every node whether it is a discrete node (sets $P^D$ and $T^D$) or a continuous one (sets $P^C$ and $T^C$);

$\text{Pre}: P \times T \rightarrow Q_+$ (for $P_i \in P^C$) or $\mathcal{N}$ (for $P_i \in P^D$) is the input incidence application;

$\text{Post}: P \times T \rightarrow Q_+$ or $\mathcal{N}$ is the output incidence application with same particularities as previous

$\mathbf{m}_0: P \rightarrow R_+$ or $\mathcal{N}$ is the initial marking.

If $P_i$ and $T_j$ are such that $P_i \in P^D$ and $T_j \in T^C$, then $\text{Pre}(P_i, T_j) = \text{Post}(P_i, T_j)$ must be verified.

A *discrete transition* in a HPN is *enabled* if each place $P_i \in {}^\circ T_j$ fulfils the condition (as for a discrete PN): $m(P_i) \geq \text{Pre}(P_i, T_j)$. A *continuous transition* in a HPN is *enabled* if the marking $m(P_i)$ of each place $P_i \in {}^\circ T_j$ meets the following conditions:

    a. $m(P_i) \geq \text{Pre}(P_i, T_j)$, if $P_i \in P^D$

    b. $m(P_i)$, if $P_i \in P^C$

The *enabling degree* of a continuous transition in HPN is similar to the case of CPN.

A *macro-marking* for a HPN is a set of markings

$$\mathbf{m}^* = (\mathbf{m}^D, \mathbf{m}^{C*}) \qquad (7)$$

such that:

1) the partial marking $\mathbf{m}^D$ is either a *marking of the discrete part*, or a *macro-marking of the discrete part* in case of unbounded discrete part

2) the partial marking $\mathbf{m}^{C*}$ is a *macro-marking of the continuous part*

In a HPN, a change of macro-marking can occur if an *event* belonging to one of the following types occurs.

*C1-event:* the marking $m(P_i)$ of $P_i \in P^C$ becomes zero.

*C2-event*: an unmarked place $P_i \in P^C$ becomes marked.

*D1-event*: firing of a transition $T_j \in T^D$.

*D2-event*: enabling degree of a transition $T_j \in T^D$ changes because of the marking of a place $P_i \in P^C$.

A *timed* HPN (THPN) is a pair
$\text{THPN} = \langle R, temp \rangle$ such that:

$R$ is a marked autonomous HPN;

*temp* is a function $temp: T \rightarrow Q_+ \cup \{\infty\}$, with

$$temp(T_j) = \begin{cases} d_j, & \text{if } T_j \in T^D \\ \dfrac{1}{U_j}, & \text{if } T_j \in T^C \end{cases} \qquad (8)$$

where $d_j$ is the timing associated with discrete transitions $T_j \in T^D$; and $U_j$ is the flow rate associated with continuous transitions $T_j \in T^C$.

The *D-enabling degree* of a transition $T_j \in T^C$ for a marking $\mathbf{m}$, denoted by $D(T_j, \mathbf{m})$, is the enabling degree of $T_j$ after *all the arcs from a place $P_i \in P^C$ to a transition $T_j \in T^C$ have been deleted*. Then,

$$D(T_j, \mathbf{m}) = \min_{P_i \in {}^\circ T_j \cap P^D} \frac{m_i}{\text{Pre}(P_i, T_j)} \qquad (9)$$

In a THPN:

a) The *flow rate* $U_j$ associated with transition $T_j$ corresponds to its maximal speed if its D-enabling degree is 1.

b) The *maximal firing speed* $V_j$ of transition $T_j$ is

$$V_j = D(T_j, \mathbf{m}) \cdot U_j \qquad (10)$$

Considering $\mathbf{m} = (\mathbf{m}^D, \mathbf{m}^C)$, where $\mathbf{m}^D$ and $\mathbf{m}^C$ are the marking associated to discrete or continuous places, the marking used for computation of speed vector is denoted by $\tilde{\mathbf{m}} = (\mathbf{m}^D, \tilde{\mathbf{m}}^C)$, where the marking $\tilde{m}_i(P_i) = \{0, 0^+, R_+\}$ for all $P_i \in P^C$.

In a THPN the *non-immediate* transition is defined as in TCPN. A non-immediate continuous transition $T_j \in T^C$ in a THPN is *enabled* at moment $t$ if $D(T_j, \mathbf{m}(t)) > 0$, and $\tilde{m}_i(t) > 0$ for every $P_i \in {}^\circ T_j \cap P^C$. If $T_j$ is enabled, it is *strongly enabled* if $m_i(t) > 0$ for every $P_i \in {}^\circ T_j \cap P^C$, and *weakly enabled* otherwise. An *immediate continuous transition* $T_j \in T^C$ in a THPN is weakly *enabled* at $t$ if $D(T_j, \mathbf{m}(t)) > 0$, and all the places $P_i \in {}^\circ T_j \cap P^C$ with $m_i(t) = 0$ are characterized by a positive feeding speed $I_i(t) > 0$.

As in the case of TCPN, the evolution graph of a THPN is built by using the concept of phase. A phase is defined such that
1. the marking $\mathbf{m}^D$ of places $P_i \in P^D$ is *constant*;
2. the enabling vector $\mathbf{e}^D$ of transition $T_j \in T^D$ is *constant*;
3. the instantaneous speed vector $\mathbf{v}$ for transitions $T_j \in T^C$ is *constant*.
4. the marking $\mathbf{m}^C$ of places $P_i \in P^C$ is constant

The vector ($\mathbf{m}^D, \mathbf{e}^D, \mathbf{v}$) define a phase in the evolution graph.

In THPN, a *change of phase* can occur only if an *event* belonging to one of the following types occurs.

*C1-event*: the marking of a marked $P_i \in P^C$ becomes zero.

*D1-event*: *firing* of a transition $T_j \in T^D$.

*D2-event*: enabling degree of a transition $T_j \in T^D$ changes because of the marking of a place $P_i \in P^C$.

There are four cases of conflicts in the THPN, depending on the type of transitions and places involved in. Denoting by $K_c = \langle P_c, \{T_a, T_b\} \rangle$ a structural conflict, the following cases are presented

*Case 1*: $T_a, T_b \in T^D$ and $P_c \in P^C \cup P^D$

This case is similar to TPN (also for continuous or discrete place $P_c$)

*Case 2*: $T_a, T_b \in T^C$ and $P_c \in P^C$.

The conflict is solved by applying priorities or sharing rules as in the case of TCPN

*Case 3*: $T_a \in T^D$ and $T_b \in T^C$, or $T_a \in T^C$ and $T_b \in T^D$

In such situation the discrete transition is always fired before the continuous transition.

*Case 4*: $T_a, T_b \in T^C$ and $P_c \in P^D$.

This case is solved by applying priorities or sharing rules. A detailed explanation is given in [4].

## 3. DSS STRUCTURE

The block architecture of the Decision Support System is presented in Figure 1. Its structure was designed taking into account the specifications of the problems it has to resolve.

The communication systems, as well as transport networks transfer a certain flow of entities (informational or physical) from one starting point to a destination, via a net of possible paths, which will be further called connections. Usually the performance evaluation criteria for transport will be the time. The availability of connections can vary, as well as the amount of entities to be transferred and the time limits. Therefore, a solution will satisfy only a given set of constraints and will be unique. An acceptable solution is searched for, not necessary the best one, even if some optimization is desirable.

Since the global network usually consists of different sub-networks, there was taken into account the possibility of composing sub-models into a global one, to be analyzed. Consequently, the DSS will include a net-editor that will allow the user to design models of networks represented in HPN formalism.

The nets can be either directly analyzed or stored in a model database. With the models from the database, the composition block permits the building of larger global nets. Finally, the history module lets the user either to store analyzed scenarios or to verify and compare previous analyses.

The user interface module allows the user to initialize the structural models with actual parameters and especially to simulate the net functioning.

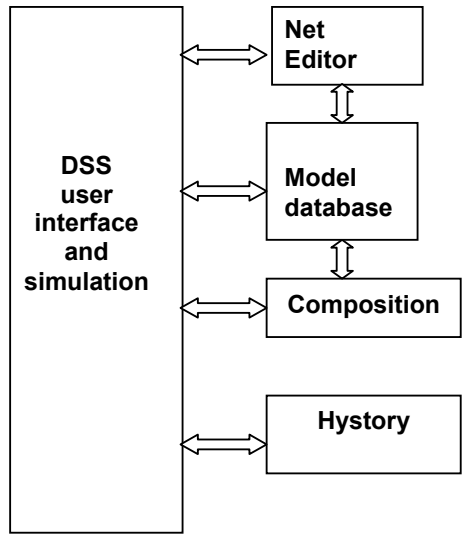

**Fig. 1** DSS structure

The following section will present the model design and analysis methodology on a small communication network – case study.

## 4. THE COMMUNICATION SYSTEM- A CASE STUDY

The considered communication system (figure 2) transfers information between two nodes (node 1 and node 4), source and destination, using any available connection, via the other network nodes (i.e., node 2 and node 3).

The information to be transferred, consisting of a given number of packets, has to reach the destination into a given amount of time. The packets can be sent by different routes, as the destination node has the possibility to order them as necessary for reconstructing the information. Therefore, an efficient distribution of packets on different paths, starting from the source node could improve the overall transmission time. The problem to be analyzed is how to distribute packets on different routes, according to their availability and transfer speed, in order to meet the overall time constraints. Obviously, the routes will not consider twice the same node.

*Hybrid Petri net model*
Figure 2 presents a communication sub-net that aims to transmit a number of information packets from the node 1 to node 4, eventually using internal nodes 2 and 3. Thus, the possible transfer routes are: 1 -> 4, 1 -> 2 -> 4; 1->3->4; 1 -> 2 -> 3 -> 4 and 1 -> 3 -> 2 -> 4 as specified in the figure 2.

As a part of a larger communication system, it is assumed that nodes 1, 2 and 3 perform also other jobs more or less important than the transfer activity. Moreover, for a particular system configuration some physical connections between nodes could be not available. The priorities of jobs and the availability of connections will be modeled by the continuous transition speeds and priorities.

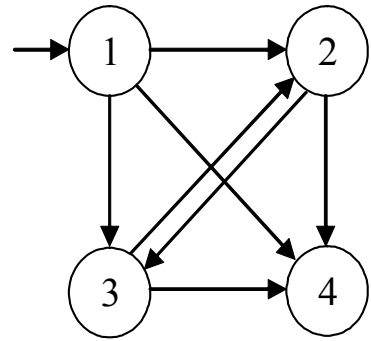

**Fig. 2**. Communication subsystem

The modeling framework is Hybrid Petri Nets mentioned in section 2, as it is suitable for the description of both continuous states or actions (transfer) and discrete states (the existence of a connection). Figure 3 illustrates the HPN structural model of the communication sub-system. For simulation purposes the marking will be initialized in order to reflect the associated connection state. It should be noted that even if a connection is available at a certain moment, it could physically break-down. The presence/absence of a connection is modeled by the set of discrete transition - place

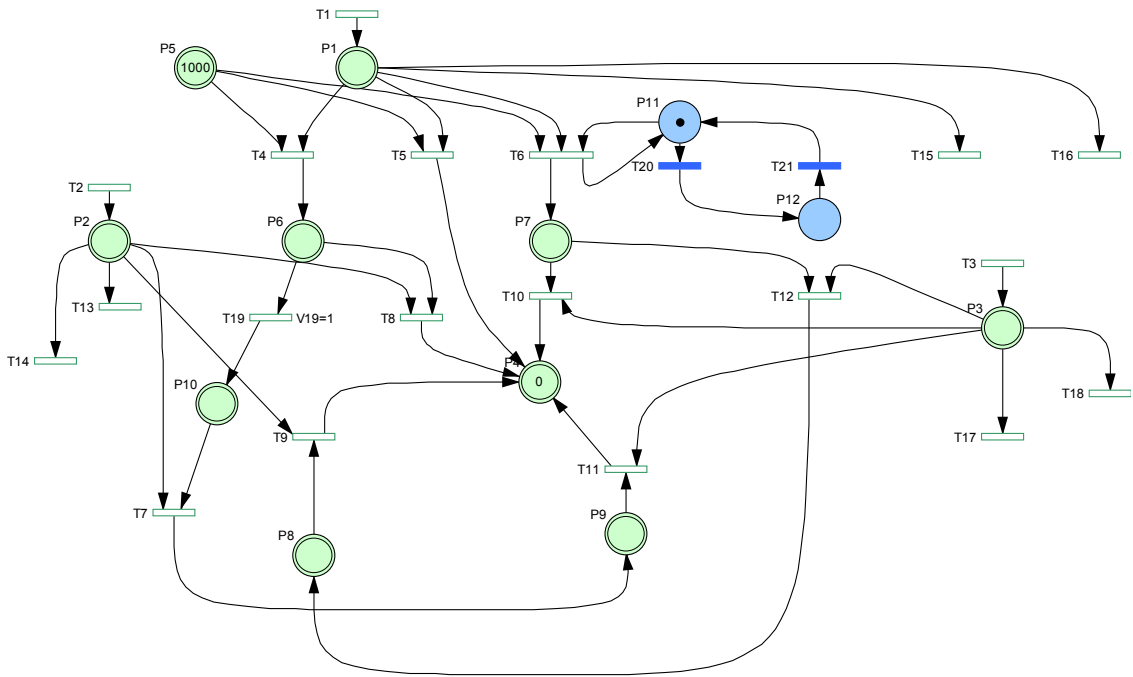

**Fig. 3.** Hybrid Petri net model

In figure 3, transitions $T_1 \div T_3$ model the working speed of the nodes $1 \div 3$. Continuous place $P_1$ is associated to node 1. The continuous transitions $T_4 \div T_6$ model the transmission speed from node 1 to nodes 2, 4 and 3 respectively. $T_{15}$ is used for modeling the other jobs the node 1 can execute, other than the transmitting activity, with higher priority. Similarly $T_{16}$ models the supplementary jobs of node 1, other than the transmitting activity, having lower priority. During the simulation of different activity scenarios, speeds and priorities can be modified in order to compare functioning regimes.

Place $P_5$ models the number of packets to be transmitted to the destination. The modeling approaches for both nodes 2 and 3 are similar: $P_2$ models node 2 and $P_3$ models node 3. From these nodes the information could be transmitted towards node 3 or 4 (from node 2), respectively nodes 2 or 4 (from node 3) by transitions $T_7$, $T_8$ and $T_9$ respectively $T_{10}$, $T_{11}$ and $T_{12}$. The transition $T_7$ models the information transmission from node 2 to node 3 and $T_{12}$ from node 3 to node 2. For the route 2 -> 4 transitions $T_8$ and $T_9$ are used that distinguish between the ways 1-> 2-> 4 and 3 -> 2 -> 4. Identically, in order to model the transmission on the route 3 -> 4 $T_{10}$ and $T_{11}$ are used (corresponding to connections 1-> 3 and 2 -> 3).

Places $P_6 \div P_9$ are intermediate and model the information flow transferring from node 1 to node 2 ($P_6$), 1 to 3 ($P_7$), 3 to 2 ($P_8$) and 2 to 3 ($P_9$). The transitions $T_{14}$ and $T_{17}$ model the tasks more important than the transmission of the nodes 2 respectively 3; the transitions $T_{13}$ si $T_{18}$ model the tasks less significant than the transmission of the nodes 2 respectively 3. The place $P_4$ corresponds to node 4. The transfer towards node 4 is already encoded in the transitions $T_5$, $T_8$, $T_9$, $T_{10}$ and $T_{11}$.

All transitions with low priority ($T_{13}$, $T_{16}$ and $T_{18}$) will have infinite maximal firing speed so that the places $P_1$, $P_2$ and $P_3$ will not accumulate tokens.

*System analysis (scenario analysis)*
Each transition $T_4 - T_{11}$ is enabled also by a discrete place. By unmarking a discrete place, the absence of the associated connection between two nodes is evidenced. There are two configuration possibilities: one of them is the setting of the discrete marking; the other consists in association of timing to discrete transitions, so as enabling /disabling (of a connection) is realized during the system analysis stage.

Different analysis scenarios could be obtained by setting the maximal speeds associated to either the transfer activities or other processing jobs. Its values could be constant a priori established, piecewise constant or stochastic (generated by the computer) on time intervals. Also by setting certain priorities/sharing for different possible transfer routes various functioning scenarios will be constructed.

In order to choose a certain route of transmission, a higher priority will be given to the modelling transition. Should no route is favorite, all the transitions will have the same weight 1 in a common sharing group. The sharing situation must be carefully chosen, as in the CPN analysis algorithms there are only three priority/sharing combinations [4].

The analysis of the communication /transport system assumes the following steps:
- setting the configuration of a communication network (the user sets the initial discrete marking that corresponds to available connections);
- setting the packet transmission speeds on each direct link. This could be done either by the user or by the system. In the last case stochastic values could be associated with the speeds.
- choosing the number of packets composing a message;
- setting the maximum time of message transmission.
Besides these initializations, before the simulation begins, a priority level has to be assigned to each transition.

*Example*
A numerical example will be analyzed by using Syrphico as simulation tool, in order to illustrate the effective speed computing. It is assumed that all connections are available. The continuous marking is set for place P5 (1000). The maximal speed vector is: $\mathbf{v}_j$=[4, 3, 5, 1, 1.5, 2, 1, 0.5, 1, 1, 1, 2, -1, 0.5, 0.5, -1, 1, -1, 1 ]. The transition priorities are illustrated in figure 4. Consequently, the instantaneous speed vector results in: $\mathbf{v}$ = [4, 3, 5, 1, 1.5, 1, 0.5, 0.5, 1, 0, 0.5, 1, 0.5, 0.5, 0.5, 0, 1, 2.5, 0.5]. The marking evolution is then: $m_1(t) = 0+$, $m_2(t) = 0$, $m_3(t) = 0$, $m_4(t) = 3.5*t$, $m_5(t) = 1000 - 3.5*t$, $m_6(t) = 0+$, $m_7(t) = 0+$, $m_8(t) = 0+$, $m_9(t) = 0+$, $m_{10}(t) = 0+$.

By inspecting the instantaneous speed vector , it could be observed that, even if a connection is available, it may not be used due to time consuming server utilization. The other parallel routes need a transmission time of 286 t.u.

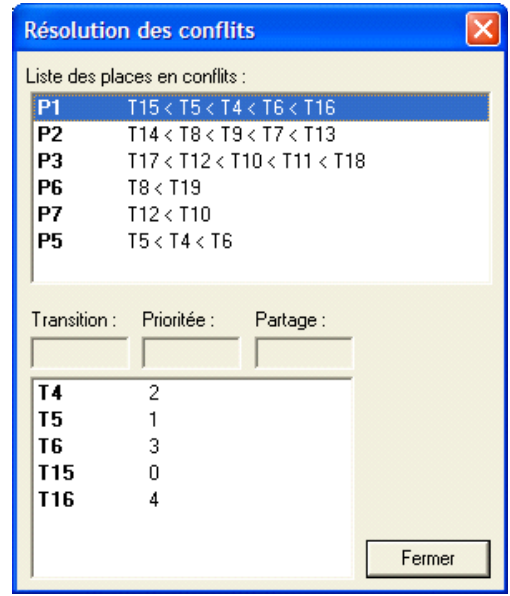

**Fig.4.** Setting the transition priorities

Taking into account two modes of solving the transition conflict, the decision support system will analyze all the possible configurations. The first configuration meeting the time criteria will be selected.

If the time criterion is not accomplished then:
- if there is a place that accumulates markings its input transition will be set with lower priority;
- a larger number of parallel routes are considered.

Before simulation starts the Decision Suport System will chose first (for transitions that model information transmission) the priority level according to the transition maximal speeds: the transition with the biggest maximal speed will have the highest priority level. But this does not always lead to best results, as shown by the following example: suppose that the Petri Net's maximal speeds are: $\mathbf{v}$ = [4, 3, 5, 3, 2, 2, 1, 0.5, 1, 1, 1, 1, ∞, 0.5, 0.5, ∞, 1, ∞]

Because $T_4$ has the biggest maximal speed (from $T_4$, $T_5$ and $T_6$), it will initially receive the highest priority level (from the three transitions).

*Case A*
Conflict resolutions are: $T_{15} < T_4 < T_5 < T_6 < T_{16}$ (for conflict resolution associated with $P_1$ place) and $T_4 < T_5 < T_6$ (for conflict resolution associated with $P_5$ place). For these maximal speeds and priority levels, the continuous Petri net will have the following evolution:

The first instantaneous transition speed vector will be: $\mathbf{v} = [4, 3, 5, 3, 0.5, 0, 0.5, 0.5, 0, 0, 0.5, 0, 1.5, 0.5, 0.5, 0, 1, 3.5, 0.5]$. Because the input transition of place $P_6$ (i.e. $T_4 - 3$) has an instantaneous speed greater than the sum of maximal speeds for output transitions of $P_6$ (i.e. $T_{19} - 0.5$ and $T_8 - 0.5$), the markings are accumulated in $P_6$. This means that:
- node 2 must have a buffer to store the parts that it cannot deliver (due to speed limitations);
- a second evolution phase is needed in order to deliver the parts from $P_6$. This evolution is characterized by the instantaneous speed vector: $\mathbf{v} = [4, 3, 5, 0, 0, 0, 0.5, 0.5, 0, 0, 0.5, 0, 1.5, 0.5, 0.5, 3.5, 1, 3.5, 0.5]$.

First evolution phase will end in 286 time units and the system will send all the packets in 857 time units (which is the end of the second evolution phase).

*Case B*
A lower priority level will be associated to transition $T_4$. Giving to $T_4$ a priority level lower then $T_5$, the conflict resolution rules for places $P_1$ and $P_5$ would be: $T_{15} < T_5 < T_4 < T_6 < T_{16}$ (for $P_1$) and $T_5 < T_4 < T_6$ (for $P_5$). For these new priority levels, the first phase of the continuous Petri net will be characterized by the following instantaneous transition speed vector: $\mathbf{v} = [4, 3, 5, 1.5, 2, 0, 0.5, 0.5, 0, 0, 0.5, 0, 1.5, 0.5, 0.5, 0, 1, 3.5, 0.5]$

The input transition of $P_6$ will still have an instantaneous speed (i.e. 1.5) bigger than the sum of maximal speeds for output transitions of $P_6$ (i.e. $0.5+0.5=1$), so a second evolution phase is needed. This is characterized by: $\mathbf{v} = [4, 3, 5, 0, 0, 0, 0.5, 0.5, 0, 0, 0.5, 0, 1.5, 0.5, 0.5, 3.5, 1, 3.5, 0.5]$

First evolution phase will still end in 286 time units and the system will send all the parts in 429 time units (which is the end of the second evolution phase).

*Case C*
Transition $T_4$ will be set with the lowest priority level. Transition $T_5$ keeps its higher priority and $T_4$ will have a priority level lower then $T_6$; the conflict resolution rules for places $P_1$ and $P_5$ would be: $T_{15} < T_5 < T_6 < T_4 < T_{16}$ (for $P_1$) and $T_5 < T_6 < T_4$ (for $P_5$). In this case the first phase of the continuous Petri net will be characterized by the following instantaneous transition speed vector: $\mathbf{v} = [4, 3, 5, 0, 2, 1.5, 0, 0, 1, 0.5, 0, 1, 1.5, 0.5, 0.5, 0, 1, 2.5, 0]$

This is the only phase needed to transfer all the packets from node 1 to node 4. The phase time is 286 time units.

From all three situations one notes that (because of node 1 limitation) only one intermediate node is selected for message transmission and the total number of routes is three.

As mentioned before, there is the possibility of increasing the number of routes. This situation appears if neither node 2 nor node 3 could send the packets as fast as they receive them.

## 5. CONCLUSION

The paper presents a decision support system constructed on a modular approach for modeling and analysis of the complex systems in communication/transport area. The partial models of intermediate nodes could be composed in order to obtain the whole system model. HPN was chosen as modeling and analysis tool due to a significant modeling power appropriate for complex systems regarded as hybrid systems.

The decision support system inspects the scenarios provided by the system analysis and proposes a time suited solution.

The future research trend is to search the optimal solution corresponding to the minimum transmission time. In this purpose, all the possible situations for priorities/sharing allocation will be analyzed.

**REFERENCES**

[1]. Alla, H. and R. David (1998). A Modelling and Analysis Tool for Discrete Event Systems: Continuous Petri Net, *Performance Evaluation*, vol. **33**, pp. 175–199.

[2]. Balduzzi, F., A. Giua, and G. Menga (2000). First-Order Petri Nets: A Model for Optimization and Control, *IEEE Trans. on Robotics and Automation*, vol. **6,** N° 4.

[3]. David, R. and H. Alla (1987). Continuous

Petri Nets, *8th European Workshop on Appli. and Theory of Petri Nets*, Saragosse, pp. 275-294.

[4]. David, R. and H. Alla (2005). *Discrete, Continuous, and Hybrid Petri Nets*, Springer, Heidelberg.

[5]. Demongodin, I., M. Mostefaoui, and N. Sauer (2000). Steady State of Continuous Neutral Weighted Marked Graphs, in *Proc. Int. Conf. on Systems, Man and Cybernetics*, IEEE / SMC, pp. 3186-3194.

[6]. Le Bail, J., H. Alla, and R. David (1991). Hybrid Petri Nets, *Proc. European Control Conference*, Grenoble.

[7]. Munteanu C.A., Stanescu A.M., Caramihai S.I., Culiţă J. (2005) Algorithmic transition grouping for speed calculation in a time continuous Petri net. Buletinul Stiintific al UPB, ISSN 1454-2331, seria C, **vol** 67, nr.4, pp. 73-80.

[8]. Recalde, L. and M. Silva (2000). PN Fluidification Revisited: Semantics and Steady State, *Proc. Int. Conf. on Automation of Mixed Processes: Hybrid Dynamic Systems* (ADPM 2000). Dortmund, pp.279-286.

[9]. Special Issue on Hybrid Petri Nets (2001). J*ournal of Discrete Event Dynamic Systems: Theory and Applications*, vol. **11**, A. Di Febbraro, A. Giua, and G. Menga Eds.